\documentclass[aps,pra,preprint,superscriptaddress,showpacs]{revtex4-1}

\usepackage{amsmath,amssymb,amstext}
\usepackage{mathrsfs}
\usepackage{dsfont}
\usepackage{graphicx}
\usepackage{color}
\usepackage{transparent}
\usepackage[all]{xy}
\usepackage{subfigure}
\usepackage[english]{babel}

\newcommand{\ZZ}{{\mathbb Z}}

\newcommand{\eps}{\varepsilon}

\newcommand{\sinc}{{\rm{sinc}}}

\newcommand{\be}{\begin{equation}}
\newcommand{\ee}{\end{equation}}
\newcommand{\bea}{\begin{eqnarray}}
\newcommand{\eea}{\end{eqnarray}}

\newcommand{\dint}[1]{\mathrm{d} #1 ~}

\begin{document}
 
\title{Quantum reflection and interference of matter waves from periodically doped surfaces}

\author{Benjamin A. Stickler}
\email{benjamin.stickler@uni-due.de}

\affiliation{Faculty of Physics, University of Duisburg - Essen, Lotharstra\ss e 1, 47048 Duisburg, Germany}

\author{Uzi Even}

\affiliation{School of Chemistry, Tel Aviv University, Tel Aviv 69978, Israel}

\author{Klaus Hornberger}
\affiliation{Faculty of Physics, University of Duisburg - Essen, Lotharstra\ss e 1, 47048 Duisburg, Germany}


\begin{abstract}
We show that periodically doped, flat surfaces can act as reflective diffraction gratings for atomic and molecular matter waves. The diffraction element is realized by exploiting that charged dopants locally suppress quantum reflection from the Casimir-Polder potential. We present a general quantum scattering theory for reflection off periodically charged surfaces and discuss the requirements for the observation of multiple diffraction peaks.
\end{abstract}

\maketitle

\section{Introduction}

Quantum reflection refers to a non-vanishing reflection probability in the absence of a classical turning point \cite{landaulifschitzIII,trostreport,friedrich02,henkel96,schoellkopf10,segev97}. It is an ubiquitous and inherently quantum mechanical phenomenon relevant for chemical reactions, for the trapping of cold atoms in optical potentials, and for the scattering of atoms off surfaces \cite{Friedman1991,Friedman95,jurisch08,shimizu01,pasquini04,pasquini06,brenig1980,brenig1982,brenig1992}.

Numerous theoretical works on quantum reflection of atoms and small particles can be found in the literature; more recent studies discuss the reflection of atoms from Casimir-Polder potentials \cite{friedrich02,cote97,eltschka2000}, of anti-hydrogen from material slabs and nanoporous materials \cite{dufour2013_1,dufour2013_2}, and of small particles from evanescent laser fields \cite{henkel96,segev97}. Further works considered the case of  temporally \cite{herwerth2013} and spatially \cite{oberst05} oscillating surfaces, and it was proposed that quantum reflection of photons from magnetic fields might even test nonlinearities of the quantum vacuum in strong fields \cite{gies2013}.

Direct experimental observations of quantum reflection include the scattering of hydrogen atoms off super-fluid Helium \cite{berkhout1989}, as well as reflection measurements of atoms from surfaces mediated by the long range Casimir potential \cite{shimizu01,druzhinina03,schoellkopf08,schoellkopf10,schoellkopf09,schoellkopf13,pasquini04,pasquini06}. The use of periodically micro-structured quantum reflection gratings enabled the detection of several diffraction orders \cite{schoellkopf08,schoellkopf13}. Such optical elements can be attractive tools for metrological applications because diffraction gratings act as mass filters and because the reflection probability from the Casimir-Polder potential depends on the particle's polarizability \cite{friedrich02}. For instance, the mass selective diffraction of a micro-structured grating allowed the direct observation of Helium trimers \cite{schoellkopf08}. The quantum reflection of Bose-Einstein condensates from a micro-structured Fresnel mirror was numerically 
studied in \cite{judd10}.

Thus far, diffraction gratings for matter waves have only been realized with the help of periodic micro-structures. These micro-structures act mainly by physically blocking a part of the incoming wave, which is a purely geometrical effect \cite{schoellkopf08}. It is a natural and open question whether quantum reflection gratings can also be realized without micro-structured surfaces. It is the aim of this paper to answer this question in the affirmative and to demonstrate that quantum reflection into different diffraction orders can be realized with flat, periodically doped surfaces, thus circumventing the requirement for micro-structures.

The paper is organized as follows: In Sec. \ref{sec:formpr} we discuss some general aspects of quantum scattering off periodic surfaces. The total interaction potential induced by periodically doped surface is then derived in Sec. \ref{sec:intpot} where we also discuss its asymptotic behavior for $y \to \infty$. In Sec. \ref{sec:refl} we show that it suffices to consider this asymptotic form and we investigate the associated specular reflection from the surface. Sec. \ref{sec:diff} then demonstrates that periodically doped surfaces can be used as diffraction gratings, and we provide our conclusions in Sec. \ref{sec:conc}.

\section{Reflection from Periodic Surface Potentials} \label{sec:formpr}

\begin{figure}
 \centering
 \includegraphics[width = 90mm]{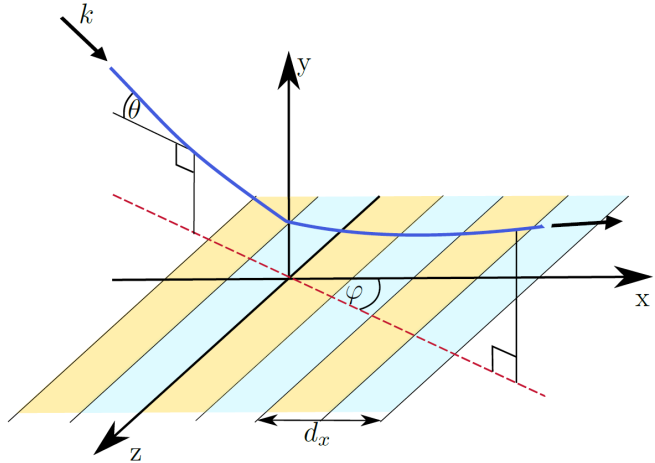}
 \caption{(Color online) Schematic of the reflection problem. The incident wavevector ${\bf k}$ impinges with angle $\theta$ at the surface, $\varphi$ is the azimuthal angle with respect to the in-plane orientation.} \label{fig:reflschem}
\end{figure}

We consider the general quantum scattering problem of a small particle of mass $m$ and static polarizability $\alpha$, off a perfectly planar but charged surface. The coordinate system is chosen as depicted in Fig. \ref{fig:reflschem}, i.e. $x$ and $z$ are the in-plane coordinates and the surface is situated at $y = 0$. The incoming beam is described by a plane-wave $\psi_{\rm in}({\bf r}) = \exp ( i {\bf k} \cdot {\bf r} )$ with $k_y < 0$. Denoting the angle of incidence by $\theta$ and the azimuthal angle by $\varphi$ we have the components of the initial wavevector $k_x = k \cos \theta \cos \varphi$, $k_y = - k \sin \theta$, and, $k_z = k \cos \theta \sin \varphi$. The angle $\theta$ is small in grazing incidence diffraction experiments and, hence, $k_x^2 + k_z^2 \gg k_y^2$, so that the energy of the lateral motion exceeds the energy of the motion towards the surface. Typically, $\theta$ is in the range $\theta \lesssim 20$ mrad \cite{schoellkopf08,schoellkopf10,schoellkopf13}.

In order to determine the reflection probability, we solve the Schr\"{o}dinger equation
\be \label{eq:statschr}
\left ( \Delta + k^2 \right ) \Psi({\bf r} ) - \frac{2 m}{\hbar^2} V({\bf r}) \Psi({\bf r}) = 0,
\ee
for the stationary scattering wave $\Psi({\bf r})$ under appropriate boundary conditions as specified below. Here, $V({\bf r})$ denotes the total interaction potential between the point particle and the surface; it will be discussed in detail in Sec. \ref{sec:intpot}.

If $V({\bf r})$ is everywhere attractive no classical turning point exists, although a finite reflectivity due to quantum reflection might occur.  The concept of Wentzel-Kramers-Brioullin (WKB) wave functions proved to be particularly powerful for the theoretical description of quantum reflection \cite{trostreport}. This is due to the fact that the validity of the WKB approximation requires that the local de Broglie wave length varies sufficiently slowly \cite{trostreport,cote97,friedrich02}, while a strong local variation of the de Broglie wave length indicates regions where quantum reflection is likely to occur. WKB waves are therefore the natural in- and out-going waves for quantum reflection problems. It is a remarkable general result that the WKB wave functions become exact both for large and small separations, $y \to \infty$ and $y \to 0$, if the interaction potential between the surface and the impinging particle vanishes faster than $1 / y^2$ for large separations $y \to \infty$ \cite{trostreport}. 
The reflection probability then approaches unity in the limit of small normal velocities $k_y \to 0$. Moreover, if $V({\bf r})$ is periodic, reflection into different diffraction orders is possible, as was observed experimentally with micro-structured reflection gratings \cite{schoellkopf08}.

\subsection{Boundary Conditions}

We proceed to specify the boundary conditions associated with Eq.~\eqref{eq:statschr} for $y \to \infty$ and $y \to 0$. For the time being, we restrict the discussion to potentials that vary periodically only in one in-plane direction, i.e. $V({\bf r}) \equiv V(x,y) = V(x+a_x,y)$. This is done for the sake of simplicity; the generalization to potentials that are also periodic in the $z$-direction is straightforward and will be briefly discussed at the end of this section.

Since the potential $V(x,y)$ is $x$-periodic with period $a_x$ and independent of $z$ the $x$-component of the outgoing wavevector can differ only by integer multiples of the grating wavevector $q_x = 2 \pi / a_x$, while the $z$-component is preserved. The $y$-component of the final momentum is determined by the conservation of total energy. Hence, the scattering state $\Psi({\bf r})$ has the general asymptotic form
\be \label{eq:bc1}
\Psi({\bf r}) \stackrel{y \to \infty}{\longrightarrow} e^{i {\bf k} \cdot {\bf r}} + e^{ik_z z} \sum_{n \in \ZZ} r_{n} e^{i ( k_x + n q_x) x}  e^{i k_{n} y},
\ee
with $k_{n} = \sqrt{k^2 - (k_x + n q_x)^2 - k_z^2}$ (which implies $k_0 = -k_y$). The reflection probability $R_n$ into diffraction order $n$ is given by $R_n = \vert r_n \vert^2$ and the total reflectivity is $R = \sum_n R_n$.

The second boundary condition, for $y \to 0$, depends on the specific behavior of the interaction potential $V({\bf r})$ near the surface $y \to 0$. If $V({\bf r})$ is proportional to $C_\gamma / y^\gamma$ in this limit, independent of $x$ and $z$, and if $\gamma > 2$, the WKB wave functions are exact near the surface \cite{trostreport}. It will become clear in Sec. \ref{sec:intpot} that the potential induced by a periodically doped surface meets this requirement due to the Casimir-Polder interaction, so that the second boundary condition can be stated as
\be \label{eq:bc2}
\Psi({\bf r}) \stackrel{y \to 0}{\longrightarrow} e^{ik_z z} \sum_{n \in \ZZ} t_n e^{i ( k_x + n q_x)x} \psi^{(n)}_{{\rm WKB}} (y).
\ee
Here, the complex numbers $t_n$ are transmission coefficients and the outgoing WKB wave $\psi_{{\rm WKB}}^{(n)}(y)$ of order $n$ is given by \cite{trostreport}
\be
\psi^{(n)}_{{\rm WKB}}(y) = \frac{1}{\sqrt{p_n(y)}} \exp \left [ - \frac{i}{\hbar} \int^y_{y_{\rm s}} \dint{y'} p_n(y') \right ],
\ee
where $p_n(y) = \sqrt{\hbar^2 k_n^2 - 2 m C_\gamma / y^\gamma}$, is the local momentum for $y \to 0$, and $y_{\rm s}$ is some arbitrary starting point.

We remark that one can estimate whether or not the WKB wave function is a good approximation to the exact solution of the Schr\"{o}dinger equation~\eqref{eq:statschr} by means of the badlands function \cite{trostreport}. In Sec. \ref{sec:refl} we will use the badlands function to estimate the reflection distance from the surface.

The stationary Schr\"{o}dinger equation \eqref{eq:statschr} together with the boundary conditions \eqref{eq:bc1} and \eqref{eq:bc2} states a well-posed boundary value problem which must be solved in order to obtain the reflection probabilities $R_n$. A more convenient formulation of the problem can be obtained by exploiting the periodicity of $V(x,y)$ in the in-plane direction $x$ directly in the Schr\"{o}dinger equation~\eqref{eq:statschr}. The resulting equations, referred to as \emph{coupled channel equations}, replace Eq.~\eqref{eq:statschr} and will be derived next.

\subsection{The Coupled Channel Equations}

Since we assume for the moment that the interaction potential $V(x,y)$ is independent of the second in-plane coordinate $z$, the component $k_z$ is preserved and the stationary scattering wave $\Psi({\bf r})$ can be expanded as
\be\label{eq:psiexp}
\Psi({\bf r}) = e^{i  k_z z} \sum_{n \in \ZZ} e^{i (k_x + n q_x ) x} \psi_n(y).
\ee
Inserting this together with the Fourier decomposition $V(x,y) = \sum_{n} V_{n}(y) e^{in q_x x}$ into the Schr\"{o}dinger equation \eqref{eq:statschr} yields the coupled channel equations \cite{manson92},
\be \label{eq:couplchan}
\left ( \partial_y^2 + k_{n}^2 \right ) \psi_{n}(y) - \frac{2 m}{\hbar^2} \sum_{n' \in \ZZ} V_{n - n'}(y) \psi_{n'}(y) = 0.
\ee
These equations describe the coupling between different diffraction orders $n$ with wavenumbers $k_n = \sqrt{k^2 - (k_x + n q_x )^2 - k_z^2}$, which are linked by the Fourier coefficients $V_n(y)$ of the potential $V(x,y)$. Comparison with Eqs. \eqref{eq:bc1} and \eqref{eq:bc2} reveals that the boundary conditions to Eq.~\eqref{eq:couplchan} are
\be \label{eq:bc3}
\psi_{n}(y) \stackrel{y \to \infty}{\longrightarrow} \delta_{n0} e^{- i k_0 y} + r_{n} e^{i k_n y},
\ee
and
\be \label{eq:bc4}
\psi_n(y) \stackrel{y \to 0}{\longrightarrow} t_n \psi_{\rm WKB}^{(n)}(y).
\ee

The coupled channel equations \eqref{eq:couplchan} in combination with the boundary conditions \eqref{eq:bc3} and \eqref{eq:bc4} are the basis for the treatment of quantum scattering and of quantum reflection off periodic surfaces. For a given interaction potential $V(x,y)$ one must in general resort to numerical methods. An algorithm for this task is presented in the appendix.

Let us fix some notations which will be helpful when discussing general potentials. The diffraction order $n = 0$ is referred to as \emph{specular}. Consequently, we refer to $V_0(y)$ as the \emph{specular potential} while $V_n(y)$, $n \neq 0$, is the $n$-th \emph{coupling potential}. Moreover, the wavenumber of the $n$-th diffraction order $k_n$ characterizes whether this diffraction order is \emph{open} ($k_n^2 > 0$) or \emph{closed} ($k_n^2 < 0$), in (imperfect) analogy to the notion of open and closed scattering channels in conventional scattering theory \cite{taylor}.

We remark that in practice it suffices to solve the coupled equations \eqref{eq:couplchan} only for the open diffraction orders, $k_n^2 > 0$. This is due to the fact that quantum reflection occurs far away from the surface where the influence of the closed order wave functions is negligible, see \cite{friedrich02,cote97,trostreport} as well as Sec. \ref{sec:refl}.

The generalization of the coupled channel equations \eqref{eq:couplchan} to a potential $V({\bf r})$ which is periodic also in $z$-direction is straightforward and comes without any further complications \cite{manson92}. In particular, a second Fourier index is added to all Fourier coefficients and one must replace $k_n$ with $k_{nm} = \sqrt{k^2 - (k_x + nq_x)^2 - (k_z + m q_z )^2}$. The resulting coupled channel equations are then the direct generalization of Eqs. \eqref{eq:couplchan}.

% In Sec. \ref{sec:diff} we shall explicitly distinguish between $x$- and $z$-diffraction orders, i.e. scattering with $\Delta k_x \neq 0$ or $\Delta k_z \neq 0$, respectively.

% We refrain from writing the full coupled channel equations for an $x$- and $z$-periodic potential $V({\bf r})$ because we won't need it in our discussion.

%a more detailed discussion of the total interaction potential $V({\bf r})$ reveals that it suffices to consider the $x$-specular potential $V_0(y,z)$. Fourier expansion in the $z$-coordinate then gives coupled channel equations of the form \eqref{eq:couplchan}, where $n$ labels the $z$-diffraction orders of the $x$-specular potential $V_0(y,z)$.

\section{The Interaction Potential} \label{sec:intpot}

We now discuss the total interaction potential for a periodically doped surface and derive its asymptotic behavior. We will see in Sec. \ref{sec:refl} that its asymptotic shape determines the probability for quantum reflection and is thus of major significance for the observation of different diffraction orders.

The total interaction potential between a polarizable point particle and a doped surface consists of two contributions: (i) the Casimir-Polder (CP) interaction $V_{\rm CP}(y)$, and, (ii) the electrostatic interaction $V_{\rm el}({\bf r})$ induced by the surface charge $\sigma(x,z)$. The purely attractive CP interaction $V_{\rm CP}(y)$ between an atom and a flat surface can be well described with the aid of shape functions \cite{friedrich02,trostreport}. It behaves as $V_{\rm CP}(y) \to - C_4 / y^4$ in the highly retarded limit $y \to \infty$ while it approaches $V_{\rm CP}(y) \to -C_3 / y^3$ for small separations \cite{Buhmann2007,buhmann1,buhmann2}. It is a matter of the particle and of the surface under investigation whether the highly retarded limit $y \to \infty$ suffices to describe quantum reflection \cite{trostreport}.

Thus, the total interaction potential $V({\bf r})$ can be written as
\be \label{eq:intpot1}
V({\bf r}) = V_{\rm CP}(y) - \frac{\alpha}{2} \vert {\bf E} ({\bf r}) \vert^2,
\ee
where the electrostatic field ${\bf E}({\bf r})$ is determined by the surface charge distribution $\sigma (x,z)$,
\be \label{eq:efeld}
{\bf E}({\bf r}) = \frac{1}{4 \pi \eps_0} \int \dint{x'} \dint{z'} \frac{\sigma(x - x',z - z')}{\left ( x'^2 + y^2 + z'^2 \right )^{\frac{3}{2}}} \left ( \begin{array}{c}
                                                                                                                                                   x' \\
																		   y \\
																		   z'
                                                                                                                                                  \end{array} \right ).
\ee
%The last relation follows directly from Coulomb's law for the charge distribution $\rho({\bf r}) = \sigma(x,z) \delta(y)$.

The total interaction potential \eqref{eq:intpot1} is everywhere attractive, thus no turning point exists. It falls off at least as $1 / y^3$ near the surface $y \to 0$ due to the CP interaction, which justifies the use of WKB functions for small normal distances $y \to 0$, see Sec. \ref{sec:formpr} and \cite{trostreport}. However, the asymptotic behavior assumed in \eqref{eq:bc2} still requires that the potential \eqref{eq:intpot1} is independent of the in-plane coordinates $x$ and $z$ as $y \to 0$.

\subsection{The Electrostatic Interaction Potential}

Let us therefore take a look at the electrostatic contribution $V_{\rm el}({\bf r})$ to the total interaction potential \eqref{eq:intpot1} of the periodically doped surface. We exploit the periodicity of the doping by expanding the charge distribution $\sigma(x,z)$ as
\be \label{eq:fourexp}
\sigma(x,z) = \sum_{n \in \ZZ} \sigma_{n}(z) e^{i n 2 \pi x / d_x},
\ee
where $d_x$ denotes the doping period (not necessarily equal to the period $a_x$ of the potential). Naturally, we consider the case of an electrically neutral surface, $\sigma_0(z) = 0$, which avoids an infinitely extended constant electric field \cite{jackson}. Moreover, for simplicity we assume that the charge distribution is symmetric in $x$, $\sigma(-x,z) = \sigma(x,z)$ and that it varies in the second in-plane direction $z$ on a length scale much larger than the doping period $d_x$.

Inserting the Fourier decomposition \eqref{eq:fourexp} of $\sigma(x,z)$ into the expression for the electric field~\eqref{eq:efeld} and noting that the integrand is sharply peaked at $z' \simeq 0$ one obtains
\bea \label{eq:efeld2}
{\bf E} ({\bf r}) & \simeq & \frac{1}{\eps_0} \sum_{n = 1}^\infty \sigma_n(z) e^{- n \kappa_x y} \left [  \sin \left ( n \kappa_x x \right ) {\bf e}_x + \cos \left (n \kappa_x x \right ) {\bf e}_y \right ],
\eea
where we abbreviated $\kappa_x = 2 \pi / d_x$. For large normal distances $y \gg 1 / \kappa_x$, the electric field \eqref{eq:efeld2} decays exponentially on the length scale $1 / \kappa_x = d_x / 2 \pi$, which is determined by the doping period $d_x$. This is due to the fact that the doping period $d_x$ is the only length scale available in this system.

Inserting the field \eqref{eq:efeld2} into the total interaction potential~\eqref{eq:intpot1} gives the electrostatic interaction potential $V_{\rm el}({\bf r})$,
\be \label{eq:intpot2}
V_{\rm el}({\bf r}) = -\frac{\alpha}{2 \eps_0^2} \left [ \beta_0(y,z) + 2 \sum_{n = 1}^\infty \beta_n(y,z)  e^{- n \kappa_x y} \cos \left ( n \kappa_x x \right ) \right ],
\ee
where we introduced the coefficients
\be
\beta_n(y,z) = \sum_{\ell = 1}^\infty \sigma_\ell(z) \sigma_{\ell + n}(z) e^{- 2 \ell \kappa_x y}.
\ee
Importantly, in the asymptotic regime $y \gg 1 /2 \kappa_x$ also the potential \eqref{eq:intpot2} decays exponentially, however on the length scale $1 / 2 \kappa_x$, which is half the scale of the asymptotic electric field \eqref{eq:efeld2}. The asymptotic shape of the electrostatic interaction potential \eqref{eq:intpot2} is, thus, completely determined by the doping period $d_x$, which is the only characteristic length in the problem. Since the electrostatic interaction decays exponentially for large normal distances $y$, the total interaction potential \eqref{eq:intpot1} decays asymptotically as $1 / y^4$ due to the retarded CP potential and, hence, the reflectivity approaches unity for small normal velocities $k_0 \to 0$ \cite{friedrich02}.

% \subsection*{Specular and Coupling Potentials}

The Fourier coefficients of the total interaction potential \eqref{eq:intpot1}, as required in the coupled channel equations \eqref{eq:couplchan}, are easily obtained from of Eqs. \eqref{eq:intpot1} and \eqref{eq:intpot2}. The specular potential $V_0(y,z)$ reads
\be \label{eq:intpot7}
V_0(y,z) = V_{\rm CP}(y) - \frac{\alpha}{2 \eps_0^2} \sum_{\ell = 1}^\infty \sigma^2_\ell(z) e^{- 2 \ell \kappa_x y},
\ee
and the $n$-th coupling potential takes on the form
\be \label{eq:intpot8}
V_n(y,z) = -\frac{\alpha}{2 \eps_0^2} \sum_{\ell = 1}^\infty \sigma_\ell(z) \sigma_{\ell + \vert n \vert}(z) e^{- ( 2 \ell + \vert n \vert )\kappa_x y}.
\ee

% In the above calculation it was assumed that the surface is perfectly flat. If the surface were physically micro-structured in a periodic fashion, as in the experiments by Zhao et al. \cite{schoellkopf08}, this structure would have to be taken into account in an appropriate fashion. In particular, in this case the electrostatic field would be of a different form and there would be a contribution of the Casimir-Polder interaction to the coupling potential due to the surface structure. Furthermore, we remark that it is a matter of the form of $\sigma(x,z)$ whether or not $\kappa_x = 2 \pi / d_x$ is  equal to $q_x = 2 \pi / a_x$.

% \subsection*{Example}

For example, a harmonic surface charge distribution in one in-plane direction, $\sigma(x,z) \equiv {\sigma} \cos \left ( \kappa_x x\right )$ with Fourier coefficients $\sigma_\ell = \sigma \delta_{\vert \ell \vert 1}/2$, yields $V_{\rm el}({\bf r}) = - \alpha \sigma^2 \exp ( - 2 \kappa_x y)/ 8 \eps^2_0 $. This potential is independent of the in-plane coordinate $x$, i.e. it is purely specular, although the charge distribution is periodic. It is finite near the surface $y \to 0$, which implies that the CP potential dominates the total interaction potential for small $y$. Hence, the WKB wave functions $\psi^{(n)}_{\rm WKB}(y)$ for the CP potential become exact for $y \to 0$, as it was presupposed in the boundary condition \eqref{eq:bc2}.

\subsection{Asymptotic Behavior of the Specular Interaction Potential} \label{subsec:xspecpot}

It is a general result that the quantum reflection of a matter wave approaching the surface with a small velocity component in the normal direction is primarily influenced by the asymptotic tail of the attractive interaction potential \cite{trostreport,friedrich02}. It is easily verified by inspection of Eqs.~\eqref{eq:intpot7} and~\eqref{eq:intpot8} that the interaction potential \eqref{eq:intpot1} is asymptotically ($y \gg 1 / 2 \kappa_x$) given by its specular interaction potential~\eqref{eq:intpot7}. At such distances from the surface the coupling to neighboring diffraction orders \eqref{eq:intpot8} is exponentially suppressed and the $x$-specular potential $V_0(y,z)$ is in leading order given by
\be \label{eq:intpot5}
V_0(y,z) \simeq V_{\rm CP}(y) - \frac{\alpha}{2} \left ( \frac{\sigma_1(z)}{\eps_0} \right )^2 e^{- 2 \kappa_x  y}.
\ee
As noted in the previous section, the asymptotic electrostatic potential \eqref{eq:intpot5} decays exponentially on the length scale $1 / 2 \kappa_x = d_x / 4 \pi$ and is independent of the precise shape of the doping profile $\sigma(x,z)$ within the unit-cell in $x$-direction.

%Importantly, this means that all neutral $x$-periodic charge distributions, independent of their actual form, yield the same asymptotic behavior \eqref{eq:intpot5}.

% \subsection*{Existence of an Electrostatic Region}

We have thus seen that the CP interaction dominates the potential \eqref{eq:intpot5} both at close and far distances from the surface, $y \to 0$ and $y \to \infty$. This justifies the boundary conditions \eqref{eq:bc2} also in the presence of the electrostatic interaction. In what follows, we focus on $z$-independent charge distributions, $\sigma_1 = {\rm const}$, for convenience, before returning to the general case $\sigma_1(z)$ in Sec. \ref{sec:diff}.

If $\sigma_1$ is sufficiently large an electrostatically dominated region exists between the CP region near the surface $y \to 0$ and the CP region far above the surface $y \to \infty$. In particular, for helium atoms and a given doping period $d_x$, this intermediate region exists for $\sqrt{\sigma_1}  d_x \gtrsim e \sqrt{e_0} \sqrt[4]{3 \pi} /  2 \sqrt[4]{4 \alpha_{\rm f}}$ where $\alpha_{\rm f} = e_0^2 / 4 \pi \eps_0 \hbar c$ is the fine-structure constant. This criterion can be obtained by equating the two different contributions of the $x$-specular potential \eqref{eq:intpot5} and minimizing in $y$. It must be emphasized that the criterion depends only on $\sqrt{\sigma_1} d_x$. For instance, for a doping period $d_x = 100$ nm, the required area density of carriers is roughly $\sigma_1 / e_0 \gtrsim 5 \times 10^{15}$ electrons / m$^{2}$.

% If $\sigma_1$ is sufficiently large an intermediate region exists, where the specular interaction \eqref{eq:intpot5} is dominated by the electrostatic contribution.
% It is briefly demonstrated in App. \ref{app:extent} that also the extent of the electrostatic region is completely determined by $\sqrt{\nu} d_x$.

% In Fig. \ref{fig:intpot1} we depict the form of $V_0(y)$ for $\nu = 10^{14}$ m$^{-2}$ and $d_x = 500$ nm ($\sqrt{\nu} d_x = 5$) as well as for  $\nu = 10^{16}$ m$^{-2}$ and $d_x = 500$ nm ($\sqrt{\nu} d_x = 50$). While in the second case an electrostatic region exists, the potential is Casimir dominated in the first case for all $y$ because the criterion \eqref{eq:charcar} is not fulfilled.
% 
% \begin{figure}
%  \subfigure[]{
%  \includegraphics[width = 70mm]{intpot4.eps}
%  }
%  \subfigure[]{
%  \includegraphics[width = 70mm]{intpot3.eps}
%  }
%  \caption{The $x$-specular interaction potential $V_0({\bf r})$, Eq.~\eqref{eq:intpot5}, (red solid line) as well as the bare Casimir-van der Waals interaction (blue dashed line) and the pure electrostatic interaction (black dash-dotted line) for, $d_x = 500$ nm and (a), $\sqrt{\nu} d_x \simeq 5$ and, (b), $\sqrt{\nu} d_x \simeq 50$.} \label{fig:intpot1}
% \end{figure}

% The question arises whether or not the specular approximation is a well justified. The answer to this question can be given by estimating the reflection distance from the surface with the aid of the badlands function \cite{trostreport}.

\section{Reflection Distance and Specular Quantum Reflection} \label{sec:refl}

%We already mentioned in Sec. \ref{sec:formpr} the badlands function. 

Within this section we first employ the maximum of the badlands function to estimate the reflection distance from the surface. We then demonstrate that periodically charged surfaces can suppress quantum reflection, by calculating the reflection probabilities off a surface that is periodically doped in one in-plane direction. The reflection distance from the surface will justify the asymptotic approximation discussed in Sec. \ref{sec:intpot}.

\subsection{Reflection Distance from the Surface} \label{subsec:badlands}

The badlands function $B(y)$ is an important tool for working with semiclassical wave functions since it indicates regions where the WKB approximation is a good estimate to the exact solution of the Schr\"{o}dinger equation \cite{trostreport,friedrich02,eltschka2000,cote97,friedrich04}. In particular the WKB waves well approximate the exact wave function, if the badlands function is small, $B(y) \ll 1$, while intervals where $B(y)$ is significantly non-zero can be regarded as regions where quantum reflection can occur \cite{friedrich04,trostreport}. Since no classical turning point exists in the potential \eqref{eq:intpot1} we use the maximum of the badlands function $B(y)$ to estimate the distance scale of quantum reflection.

The badlands function $B(y)$ is defined as \cite{trostreport}
\be
B(y) = \hbar^2 \left \vert \frac{3}{4} \frac{[p'(y)]^2}{[p(y)]^4} - \frac{p''(y)}{2 [p(y)]^3} \right \vert,
\ee
where $p(y)$ is the local $x$-specular momentum (we still assume $z$-independence in this section), i.e. $p(y) = \sqrt{\hbar^2 k_0^2 - 2m V_0(y)}$, with $V_0(y)$ the $x$-specular potential \eqref{eq:intpot5}; the primes denote derivatives with respect to $y$.

If the total interaction potential is dominated by the CP interaction at all distances $y$ we may neglect the electrostatic contribution to the total potential \eqref{eq:intpot5}. To keep the argument as simple as possible, we restrict the following discussion  to the case that the highly retarded limit $y \to \infty$ of the CP interaction is sufficient to describe quantum reflection, as it is the case e.g. for helium atoms reflected from silica surfaces \cite{trostreport}. In this limit the CP potential is of the form $V_{\rm CP}(y) = - C_4 / y^4$ with $C_4 = 3 \hbar c \alpha / 32 \pi^2 \eps_0$ \cite{Buhmann2007} and the maximum of the badlands function is approximately at
\be \label{eq:ymc}
y_{\rm CP} = \frac{1}{\sqrt{v_0 \sin \theta}}\sqrt[4]{\frac{3 \hbar c}{16 \pi^2 \eps_0} \frac{\alpha}{m}},
\ee
above the surface. The scale of the reflection distance is thus in this case determined completely by the properties of the particle, and independent of the doping period $d_x$ and charge density $\sigma_1$. Moreover, since the reflection distance depends only on the ratio between polarizability and mass, $\alpha / m$, it is approximately the same for atomic clusters of arbitrary size. However, the reflection probability decreases with increasing cluster size \cite{trostreport}.

On the other hand, if the electrostatic contribution dominates the quantum reflection, the maximum of the badlands function lies approximately at
\be \label{eq:yme}
y_{\rm el} = \frac{d_x}{4 \pi} \ln \left [ \frac{64 \pi^3 \alpha_{\rm f}}{3 \left ( 5 - \sqrt{21} \right )} \frac{\sigma_1^2}{e_0^2} y^4_{\rm CP}\right ],
\ee
above the surface. The reflection distance then increases linearly with the doping period $d_x$ and depends only logarithmically on the charge density $\sigma_1$ and the CP reflection length scale $y_{\rm CP}$.

For example, helium atoms approaching the surface with an initial velocity of $v_0\simeq 300$ m s$^{-1}$ at an incidence angle $\theta \geq 1$ mrad, have a CP reflection distance of $y_{\rm CP} \lesssim 40$ nm above the surface.  For an exemplary charge carrier density $\sigma_1 /e_0 \simeq 10^{16}$ electrons / m$^{2}$ the maximum of the badlands function \eqref{eq:yme} is then at a position $y_{\rm el} > 1 / 2 \kappa_x = d_x / 4 \pi$ which is deep in the asymptotic regime of the electrostatic interaction potential \eqref{eq:intpot2}. This justifies the specular approximation of the total interaction potential \eqref{eq:intpot5} for this charge density.

% of the above the surface, which is well in the asymptotic regime of
% implies that the reflection distance from the surface $y_{\rm em}$ 
% where the contribution of the coupling potentials \eqref{eq:intpot8} to the total interaction potential \eqref{eq:intpot1} is significant, $y < 1/ 2 \kappa_x$,
% The reflection distance \eqref{eq:yme} can be shifted towards the surface, out of the asymptotic region $y \geq 1 / 2 \kappa_x = d_x/4 \pi$,  only by decreasing the surface charge density $\sigma_1$. In this case the reflectivity tends to zero.

In fact, it can be argued that the asymptotic approximation \eqref{eq:intpot5} is sufficient to describe quantum reflection for arbitrary charge densities $\sigma_1$. For large densities it is valid due to the mentioned large reflection distances \eqref{eq:yme}, while for small densities the reflectivity from the purely electrostatic interaction potential tends towards zero (for the exact and for the asymptotic electrostatic interaction). The physical reason for the latter is that in case of small doping periods $d_x$ the electric field can be neglected because the CP potential dominates the total interaction potential \eqref{eq:intpot5}; on the other hand, in case of large $d_x$ the reflection probability approaches zero as will be shown in the next section.

\subsection{Specular Quantum Reflection} \label{subsec:specrefl}

We now explicitly calculate the probability for quantum reflection from a flat surface that is periodically charged with period $d_x$ in one in-plane direction. This requires solving the Schr\"{o}dinger equation \eqref{eq:statschr} with the asymptotic interaction potential \eqref{eq:intpot5} for $\sigma_1 =$ const. It can be seen from the coupled channel equations \eqref{eq:couplchan} that this is equivalent to solving these equations only for the isolated specular channel,
\be \label{eq:speceq}
\left ( \partial_y^2 + k_0^2 \right ) \psi_0(y) - \frac{2 m}{\hbar^2} V_0(y) \psi_0(y) = 0,
\ee
with the specular boundary conditions \eqref{eq:bc3} and \eqref{eq:bc4} ($n = 0$).

If the interaction potential is dominated by the CP interaction for all normal distances $y$, the reflection probability can be obtained in leading order from zero energy solutions of Eq.~\eqref{eq:speceq} \cite{cote97}. It then tends towards unity for decreasing normal velocity $k_0 \to 0$ and decreases, in the highly retarded limit, for increasing values of the characteristic CP length $b_{\rm CP} = \sqrt{2 m C_4} / \hbar$ \cite{trostreport}.

On the other hand, if the interaction potential has a dominantly electrostatic region, Eq.~\eqref{eq:couplchan} describes reflection from an exponential quantum well and can be solved analytically (since substitution $\xi \propto e^{ - \kappa_x y}$ turns Eq.~\eqref{eq:speceq} into a Bessel differential equation of complex-order) \cite{landaulifschitzIII,henkel96}. In this case the total reflection probability $R$ is \cite{henkel96}
\be \label{eq:totref}
R = e^{- k_0 d_x}.
\ee

The reflection probability tends towards unity for vanishing normal velocities $k_0 \to 0$ and decreases with increasing doping period $d_x$. The physical reason is that for a given doping period $d_x$ the asymptotic electric field \eqref{eq:efeld2} and therefore also the asymptotic interaction potential \eqref{eq:intpot2} decays exponentially on the scale defined by the period $d_x$. In the limit of large $d_x$ the asymptotic electrostatic potential \eqref{eq:intpot5} decays slowly and thus the probability for quantum reflection tends to zero.

% If $d_x$ is large compared to the length scale $b_{\rm CP}$ of the CP reflection, the interaction potential is, compared to the same scale, slowly varying and, thus, the probability for quantum reflection is low. On the other hand, if the period $d_x$ is small, high doping concentrations $\sigma_1$ are required to sustain a dominantly electrostatic region, Sec. \ref{sec:intpot}. In contrast to the CP length $b_{\rm CP}$, the length scale of the electrostatic interaction is independent of the particle's polarizability and the amplitude of the electrostatic interaction potential \eqref{eq:intpot5} 
% determines only the reflection distance from the surface \eqref{eq:yme}.

Returning to the discussion of the length scale \eqref{eq:yme} of quantum reflection, we now see that it is indeed hardly possible to observe different diffraction orders if the charge density is periodic only in $x$-direction. The reason is that for a significant coupling between the $x$-channels the surface charge has to be small because of Eq. \eqref{eq:yme}. This in turn requires large doping periods $d_x$ implying low reflection probabilities \eqref{eq:totref}. However, one can create grating structures by locally suppressing quantum reflection in a controlled fashion, as we discuss next.

\section{Quantum Reflection into Different Diffraction Orders} \label{sec:diff}

In the previous section we saw that if the surface charge $\sigma(x,z)$ is periodic in a single in-plane direction $x$, its period $d_x$ defines the dominant length scale of the system. This length scale then determines the probability of quantum reflection \eqref{eq:totref}, implying that the coupling to different diffraction orders is suppressed.

Thus, for observing diffraction into different diffraction orders a second length scale has to be introduced. This is most easily achieved by making the surface charge density $\sigma(x,z)$ periodic also in the second in-plane direction $z$. (Alternatively, one can think of a one-dimensional superstructure with two different periods.) We then deal with a two-dimensional periodic surface charge $\sigma(x,z)$ inducing the electric field which attracts the impinging particle. Again, the reflection probability is determined by the asymptotic shape of the electrostatic interaction potential in normal direction $y$. It follows from Eq. \eqref{eq:efeld2} that if the period $d_z$ in one in-plane direction is much greater than the period $d_x$ in the other in-plane direction, $d_z \gg d_x$, the asymptotic electric field decays on the length scale set by the smaller period $d_x$, while the $z$-dependence of the surface charge $\sigma_1(z)$ determines the amplitude of the asymptotic field \eqref{eq:efeld2}. At 
positions $z$ where the charge $\sigma_1(z)$ is large quantum reflection can be 
suppressed according to the discussion of the previous section, while at positions $z$ where $\sigma_1(z)$ is close to zero CP reflection can occur. In this way it is possible to realize a reflective diffraction grating in  direction $z$ consisting of an alternating sequence of absorptive `grating bars` and reflective regions in-between. Each absorptive `grating bar` is realized by short-period doping in the $x$-direction perpendicular to the grating direction $z$.

\subsection{Theoretical Description}

We thus consider a two-dimensional periodic surface charge density $\sigma(x,z)$ with periods $d_x$ and $d_z$, having a small ratio $d_x / d_z \ll 1$ so that the asymptotic electric field \eqref{eq:efeld2} in normal direction is determined by $d_x$. The asymptotic behavior of the total interaction potential is then given by the specular potential \eqref{eq:intpot5} where the charge modulation $\sigma_1(z)$ is now allowed to be a periodic function of the grating direction $z$. The asymptotic total potential \eqref{eq:intpot5} is independent of the in-plane coordinate $x$, making $k_x$ a preserved quantity. Moreover, the potential decays exponentially in the normal direction with the length scale set by the smaller period $d_x$. If the charge modulation $\sigma_1(z)$ varies smoothly with period $d_z$ between zero and some finite value $\sigma_{\rm m}$, one observes alternatingly regions where quantum reflection can occur from the CP potential ($\sigma_1(z) \simeq 0$) and regions where quantum reflection is 
suppressed due to the presence of the electrostatic interaction ($\sigma_1(z) \simeq \sigma_{\rm m}$).

To describe the reflection probability off this diffraction grating, we solve the stationary Schr\"{o}dinger equation \eqref{eq:statschr} for the specular potential \eqref{eq:intpot5},
\be \label{eq:schr2}
\left ( \partial_y^2 + \partial_z^2 \right ) \psi(y,z) - \frac{2 m}{\hbar^2} V_0(y,z) \psi(y,z) = 0,
\ee
where the total scattering state is $\Psi({\bf r}) = e^{i k_x x} \psi(y,z)$. Repeating the steps described in Sec. \ref{sec:formpr} yields the coupled channel equations for the diffraction in the $z$-direction in direct analogy to Eqs. \eqref{eq:couplchan}. They can be solved numerically with the algorithm explained in the appendix.

To demonstrate the working-principle of the diffraction grating it suffices to regard an initial wave packet approaching the surface along the short period in-plane direction $x$. Thus we have $\varphi = 0$ and $k_z = 0$. This means that the asymptotically outgoing diffraction orders have in-plane momentum $k_z' = n q_z$, where $q_z = 2 \pi / a_z$ denotes the lattice wavenumber in $z$-direction, while the normal component $k_y$ changes to $k_n = \sqrt{k^2 - k_x^2 - ( n q_z)^2}$ according to the conservation of energy. Since $k_z = 0$ the expected diffraction pattern is symmetric $R_n = R_{-n}$.

For long grating periods $d_z \gg d_x$ it is natural to assume that $2 \pi / d_z k_0 \ll 1$, which is usually known as the sudden approximation \cite{manson92}. This means that all outgoing waves have approximately the same wavenumber, $k_n \simeq k_0$. Then the Schr\"{o}dinger equation \eqref{eq:schr2} simplifies to
\be \label{eq:couplchan3}
\left ( \partial_y^2 + k_0^2 \right ) \psi(y,z) - \frac{2 m}{\hbar^2} V_0(y,z) \psi(y,z) = 0.
\ee
Here, the grating direction $z$ enters the equation only in a parametric fashion, and hence also the scattering state $\psi(y,z)$ depends on $z$ only parametrically. The corresponding asymptotic boundary condition \eqref{eq:bc1} can be expressed as
\be \label{eq:asymp1}
\psi(y,z) \stackrel{y \to \infty}{\longrightarrow} e^{-i k_0 y} + r(z) e^{i k_0 y}.
\ee

In general, the resulting reflection probability is a function of the grating direction $z$, $R(z) = \vert r(z) \vert^2$ and the diffraction probability into the $n$-th order $R_n = \vert r_n \vert^2$ is obtained from
\be \label{eq:fourierdiff}
r_n = \frac{1}{a_z} \int_{-\frac{a_z}{2}}^{\frac{a_z}{2}} \dint{z} r(z) e^{- i n q_z z}.
\ee
It is easily verified that $R = (1 / a_z) \int_{-a_z / 2}^{a_z / 2} \dint{z} R(z)$ is the total reflectivity.

In order to discuss particular examples we rewrite the Schr\"{o}dinger equation~\eqref{eq:couplchan3} by defining the dimensionless periodic function $\Delta(z)$ by $\sigma_1(z) = \sigma_{\rm m} \Delta(z)$ where $\sigma_{\rm m} = {\rm const}$ is the doping amplitude. Hence, the $n$-th order coupling potential $V_n(y)$ between diffraction orders in the grating direction $z$ is given by
\be \label{eq:coupl2}
V_n(y) = - \frac{\alpha}{2 a_z} \left ( \frac{\sigma_{\rm m} e^{-\kappa_x y}}{\eps_0} \right )^2 \int_{- \frac{a_z}{2}}^{\frac{a_z}{2}} \dint{z} \Delta^2(z) e^{- i n q_z z}.
\ee

% we rewrite Eq.~\eqref{eq:couplchan3} as
% \be \label{eq:couplchan4}
% \left ( \partial_y^2 + k_0^2 \right ) \psi(y,z) - \frac{2 m}{\hbar^2} V_{\rm CP}(y) \psi(y,z) - \frac{2 m}{\hbar^2} \Delta^2(z) \overline{V}_{\rm el}(y) \psi(y,z) = 0,
% \ee
% where we abbreviated the $z$-independent electrostatic interaction
% \be
% \overline{V}_{\rm el}(y) = - \frac{\alpha}{2} \left ( \frac{\sigma}{\eps_0} \right )^2 e^{- 2 \kappa_x y}.
% \ee

% the coupling potentials $V_n(y)$ between $z$-diffraction orders are given by
% \be \label{eq:coupl2}
% V_n(y) = \frac{\overline{V}_{\rm el}(y)}{d_z} \int_{- \frac{d_z}{2}}^{\frac{d_z}{2}} \dint{z'} \Delta^2(z) e^{- i n \kappa_z z},
% \ee
% where $\kappa_z = 2 \pi / d_z$. Before presenting our numerical results, we provide an analytic approximate solution of Eq.~\eqref{eq:couplchan4} for a specific situation.

\subsection{Stripe Geometry}

% \begin{figure}
%  \centering
%  \subfigure[]{
%  \includegraphics[width = 70mm]{refldiff1.eps}
%  }
% 
%  \caption{Reflection probabilities $R_n$ into different diffraction orders for the idealized case that the surface is doped in stripes, $f = 1 / 2$ in \eqref{eq:delta1}. We show the numerically obtained values (red bars / squares) in comparison with the analytic result \eqref{eq:refdiff} (blue dots), where $R_{\rm CP}$ has been calculated numerically. In (a) we show the diffraction peaks while in (b) we show the same reflection probabilities on a logarithmic scale.} \label{fig:diff1}
% \end{figure}

As a first example we discuss the case that the `grating bars` are of the form of rectangular doping stripes in grating direction $z$,
\be \label{eq:delta1}
\Delta (z) = \sum_{n \in \ZZ} \Theta \left ( \frac{f d_z}{2} - \vert z - n d_z \vert \right ).
\ee
Here, $f \in [0,1)$ is the opening fraction of the grating. Clearly, in this case $\Delta(z)$ is not slowly varying in direction of the grating $z$ so that the validity of the asymptotic interaction potential Eq.~\eqref{eq:intpot5} is questionable; however, we ignore this for the moment.

We solve the Schr\"{o}dinger equation \eqref{eq:couplchan3} for all positions $z$ in the grating direction. If $z$ is within the `grating bar`, $\Delta(z) = 1$, the reflection coefficient vanishes approximately, $r(z) =0$, because quantum reflection is suppressed by the electrostatic interaction. On the other hand, at positions $z$ where $\Delta(z) = 0$, the reflection coefficient is close to the value of pure CP reflection of a flat surface, $r(z) \simeq r_{\rm CP}$. In summary, we obtain
\be
r(z) = r_{\rm CP} [1 - \Delta(z) ],
\ee
and the probability for reflection into the $n$-th diffraction order $R_n = \vert r_n \vert^2$ reads as
\be \label{eq:refdiff}
R_n = R_{\rm CP} (1 - f)^2 \sinc^2 [ n (1 - f) \pi ],
\ee
with $R_{\rm CP} = \vert r_{\rm CP} \vert^2$. This is exactly the diffraction intensity expected from a periodic grating with opening fraction $(1 - f)$ \cite{bornwolf}.

%The total reflectivity $R$ is given by $R =  (1 - f) R_{\rm CP}$.

In addition, we note that the total reflectivity $R$ is given by $R = (1 - f) R_{\rm CP}$, i.e. it depends on the properties of the impinging particle through $R_{\rm CP}$. In particular, denoting by $b_{\rm CP}$ the characteristic length of the CP potential, the CP reflection probability $R_{\rm CP}$ approaches unity for $k_0 b_{\rm CP} \to 0$ and it decreases exponentially with increasing $k_0 b_{\rm CP}$ \cite{trostreport}. If the CP length scale $b_{\rm CP}$ is dominated by the highly retarded limit, $b_{\rm CP} = \sqrt{2 m C_4} / \hbar$ \cite{trostreport}, the reflectivity decreases exponentially with $v_0 \sin \theta \sqrt{\alpha m}$.

We compare the analytic result Eq.~\eqref{eq:refdiff} with the exact, numerically obtained reflectivities for helium atoms with the initial velocity $v_0 = 300$ m s$^{-1}$, approaching the surface at incidence angle $\theta = 1$ mrad. The short doping period is $d_x = 500$ nm while the long doping period is $d_z \simeq 40$ $\mu$m; the doping amplitude is set to $\sigma_{\rm m} / e_0 = 10^{16}$ electrons / m$^{2}$ and the grating fraction is given by $f = 1 / 2$. The coupled channel equations \eqref{eq:couplchan} are solved numerically with the help of Johnson's log-derivative method, see the appendix. The coupling potentials $V_n(y)$ which appear in the coupled channel equations \eqref{eq:couplchan} are given by
\be
V_n(y) = - \frac{\alpha f}{2} \left ( \frac{\sigma_{\rm m} e^{- \kappa_x y}}{\eps_0} \right )^2 \sinc ( n f \pi) .
\ee
The probability $R_{\rm CP}$ for flat CP reflection required in \eqref{eq:refdiff} is calculated numerically with the help of the same method.

\begin{figure}
 \centering
  \subfigure{
 \includegraphics[width = 75mm]{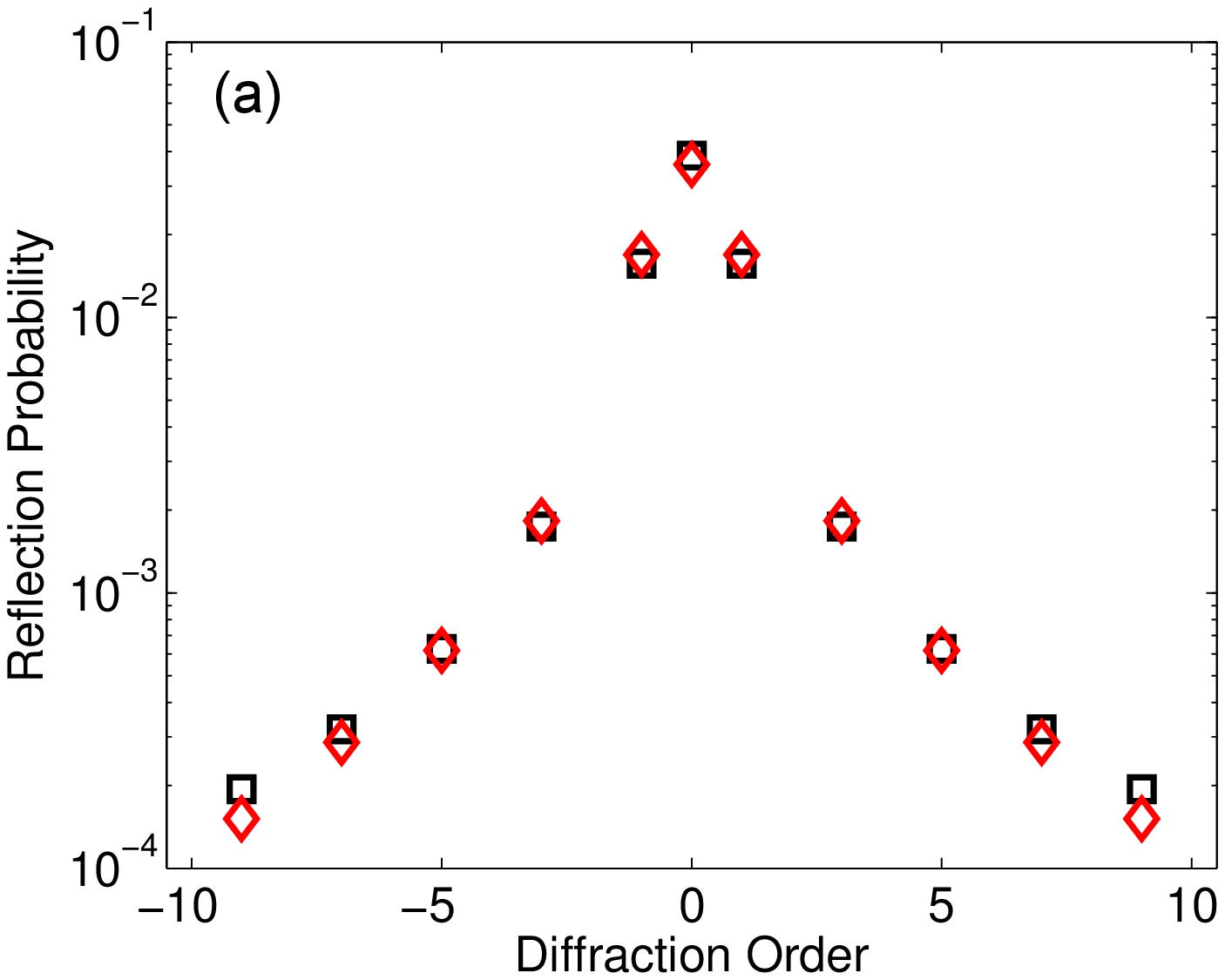}
 }
 \subfigure{
 \includegraphics[width = 75mm]{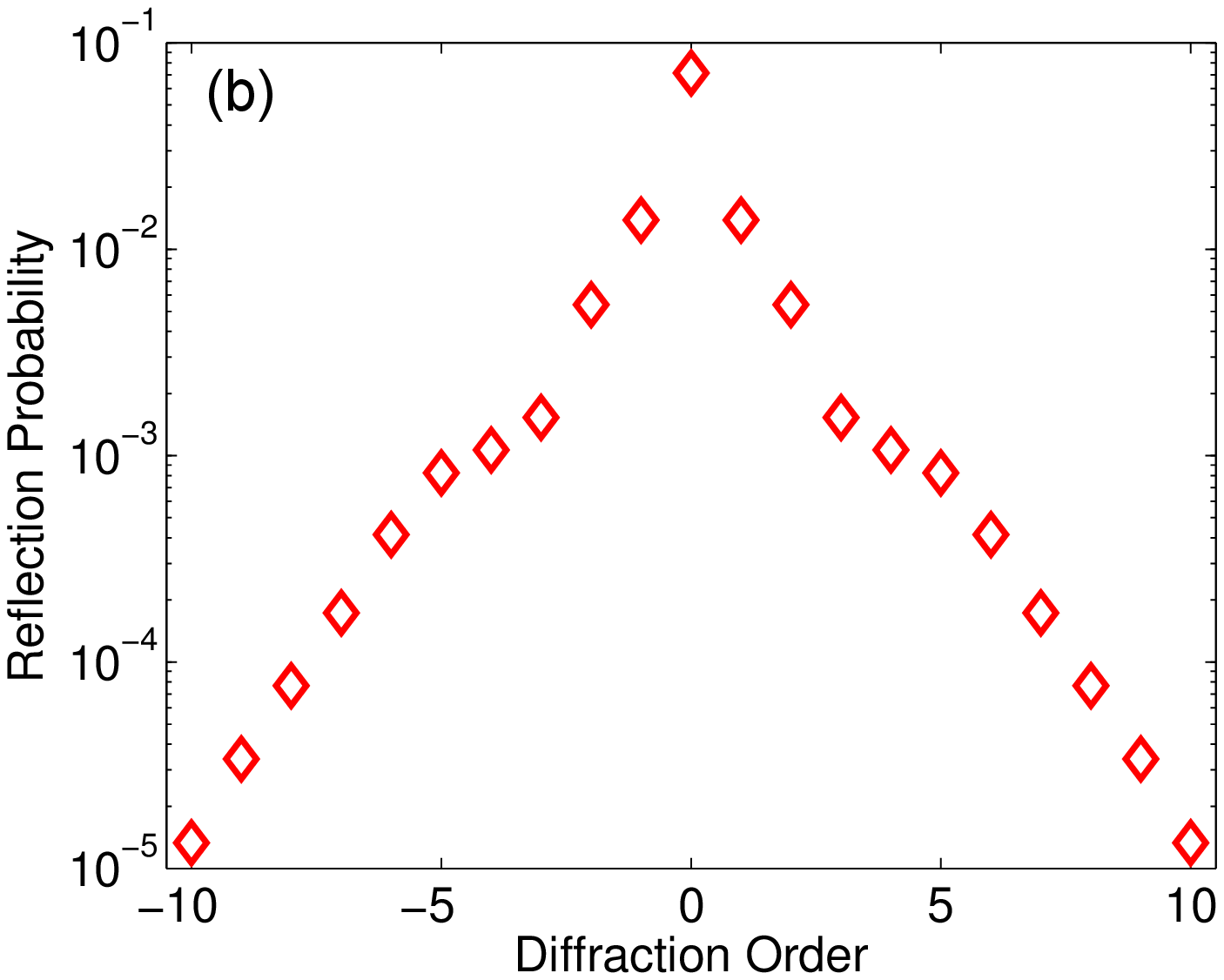}
 }
 \caption{(Color online) Reflection probabilities into different diffraction orders for a doping profile $\Delta(z)$ given by (a) Eq.~\eqref{eq:delta1} and (b) Eq.~\eqref{eq:delta3}. In (a) the numerically obtained values (red diamonds) compare well with the analytical result (black squares) given by Eq.~\eqref{eq:refdiff}.} \label{fig:diff1}
\end{figure}

In Fig. \ref{fig:diff1} (a) we see an agreement between the numerically obtained diffraction probabilities and the analytically approximated values \eqref{eq:refdiff}. In particular, while the specular peak is slightly overestimated by \eqref{eq:refdiff}, the reflection probabilities for the first diffraction orders are well described. This example clearly shows that the flat diffraction grating can be realized with the help of two-dimensional periodic surface doping.

% \subsection{Two Further Examples}

% \begin{figure}
%  \centering
%  \subfigure[]{
%  \includegraphics[width = 70mm]{refldiff2.eps}
%  }
%  \subfigure[]{
%  \includegraphics[width = 70mm]{refldiff2log2.eps}
%  }
%  \caption{Reflection probabilities $R_n$ into different diffraction orders for $\Delta(z)$ given by Eq.~\eqref{eq:delta2}. We show the numerically obtained values (red bars / squares). In (a) we show the diffraction peaks while in (b) we show the same reflection probabilities on a logarithmic scale.} \label{fig:diff2}
% \end{figure}

% We consider
% \be \label{eq:delta2}
% \Delta(z) = \sigma \cos^4 ( \kappa_z z).
% \ee
% In this case the coupling potentials $V_n(y)$ are of the form
% \be
% V_{2n}(y) = \frac{315}{2} \frac{\sigma}{\Gamma \left ( 5 + n \right ) \Gamma \left ( 5 - n \right )} \overline{V}_{\rm el}(y),
% \ee
% for $\vert n \vert \leq 4$, and zero otherwise. The diffraction pattern for $v_0 = 150$ m s$^{-1}$, $\theta = 2$ mrad. $\nu = 10^{15}$ m$^{-2}$ and $d_x = 500$ nm, is shown in Fig. \ref{fig:diff2}. Again, coupling to higher diffraction orders is observable.

More complicated doping structures $\Delta(z)$ can only be treated by solving the coupled channel equations \eqref{eq:couplchan} numerically. As a second example, we investigate the diffraction pattern resulting from a Gaussian doping profile, which might be much more realistic. Specifically, the doping profile  $\Delta(z)$ is of the form,
\be \label{eq:delta3}
\Delta (z) = \sum_{k \in \ZZ} \exp \left ( - \frac{(z - k d_z)^2}{2 \eps^2} \right ),
\ee
where we assume that the width $\eps \ll d_z$ so that there is negligible overlap between neighboring `grating bars`. The coupling potentials $V_n(y)$ are then of the form
\be
V_n(y) = - \frac{\alpha \eps \sqrt{\pi}}{2d_z}\left ( \frac{\sigma_{\rm m} e^{- \kappa_x y}}{\eps_0} \right )^2 \exp \left [ - \left( \frac{n \eps \pi}{d_z} \right )^2 \right ].
\ee
The diffraction pattern is shown in Fig. \ref{fig:diff1} (b) for helium atoms. The parameters are as above, except for the doping amplitude is $\sigma_{\rm m} / e_0 = 10^{15}$ electrons / m$^{2}$; the width of the doping profile is $\eps = d_z / 10$. Again, we find that the observation of multiple diffraction peaks is clearly possible. Compared with the previous example, the reflection probability of the specular peak is enhanced, while the reflection probabilities into the first diffraction orders are of a comparable magnitude.

% \section{Experimental Realization} \label{sec:exp}
% 
% \begin{itemize}
%  \item surface preparation:
%     \begin{itemize}
%       \item waver material and dopands
%       \item Doping concentrations
%       \item grating periods
%       \item surface 'really' flat?
%     \end{itemize}
%  \item atom beam and atom detection
%     \begin{itemize}
%       \item velocities
%       \item coherence
%       \item incidence angle
%       \item required reflectivity?
%      \end{itemize}
% \end{itemize}

\section{Conclusion} \label{sec:conc}

We demonstrated that quantum reflection of atomic and molecular matter waves into multiple diffraction orders can be achieved with flat, periodically charged surfaces. The electric field generated by the surface dopants provides a force in addition to the Casimir-Polder interaction that can locally suppress quantum reflection. Diffraction into different orders is then achievable by a two-dimensional doping pattern that is periodic in both surface directions. The diffraction grating consists of an alternating sequence of absorptive grating `bars` and reflective `grating slits`, where each `grating bar` is realized by periodic surface doping. The quantum reflection occurs from the Casimir-Polder potential in the `grating slits`, i.e. in un-doped regions between two neighboring `grating bars`. The reflection probabilities are comparable in magnitude and scaling to previous diffraction experiments off the  Casimir-Polder interaction of micro-structured gratings. Mass selection and the observation of multiple 
diffraction peaks should thus be possible also with flat, doped surfaces.

In the course of this work a full quantum scattering theory of matter waves from periodically charged surfaces was developed. The investigation of quantum reflection from a surface which is periodically doped only in one in-plane direction revealed that the reflectivity can be efficiently suppressed. This is due to the fact that the asymptotic shape of the electrostatic interaction potential is fixed by the doping period $d_x$. It was argued, by estimating the reflection distance from the surface, that in this case the asymptotic interaction potential suffices completely to describe quantum reflection. Moreover, the asymptotic shape of the interaction potential was found to be independent of the specific doping structure so that the coupling to different diffraction orders is suppressed if the surface doping is periodic in only a single in-plane direction. Based on this observation we proposed realizing flat diffraction gratings for matter waves with a two-dimensional grating-like structure. We determined 
the diffraction intensities for helium atoms and demonstrated that the observation of several diffraction peaks is indeed possible. This makes the discussed set up an alternative to micro-structured matter wave diffraction elements.

It is a natural question whether the idea of modulating the reflection by periodic doping can be extended to realize further flat optical elements, such as flat concave mirrors or Fresnel mirrors, by appropriately adjusting the surface doping. Such optical elements for quantum reflection may well be easier to produce than their micro-structured counterparts since the asymptotic form of the interaction potential is independent of the precise doping structure within the absorptive `bars`.

\section{Acknowledgments}

We thank A. R. Barnea for helpful discussions, and we acknowledge support by the European Commission within NANOQUESTFIT (No. 304886).

\appendix

\section{Numerical Method} \label{app:nummeth}

To solve the coupled channel equations \eqref{eq:couplchan} (where $n$ labels $x$- or $z$-diffraction orders) we employ Johnson's log-derivative method \cite{johnson73}. It is based on formulating Eq.~\eqref{eq:couplchan} as a matrix equation for the matrix ${\bf \Psi} ({y}) = [\Psi_{nm}(y) ]$, where $\Psi_{nm}(y)$ denotes the $n$-th component of the $m$-th solution vector associated with incident momentum $k_m$. Hence, the matrix elements of interest are $\Psi_{n0}(y)$ with incident momentum $k_0$. Defining the logarithmic derivative matrix ${\bf Z}(y) = {\bf \Psi}'(y) {\bf \Psi}^{-1}(y)$ gives a matrix Ricatti equation
\be
{\bf Z}'(y) + {\bf Z}^2(y) + {\bf U}(y) = {\bf 0},
\ee
with the potential matrix ${\bf U}(y)$ defined as
\be
U_{nm}(y) = k_n^2 \delta_{nm} - \frac{2 m }{\hbar^2} V_{n - m}(y).
\ee

Since the badlands function $B(y)$ vanishes for $y \to 0$, see Sec. \ref{sec:intpot}, we know that ${\bf Z}(y)$ is in this limit well approximated by the WKB solution with $V(y) \simeq V_{\rm CP}(y)$. Therefore, ${\bf Z}(y)$ is for $y \to 0$ given by
\be \label{eq:z0}
{\bf Z}(y) \stackrel{y \to 0}{\longrightarrow} {\rm diag} \left [ -\frac{i}{\hbar} p_n(y)- \frac{p_n'(y)}{2 p_n(y)} \right ],
\ee
where $p_n(y) = \sqrt{\hbar^2 k_n^2 - 2m V_{\rm CP}(y)}$ is the local momentum of the $n$-th diffraction order. The WKB solution \eqref{eq:z0} is then propagated with the help of Johnson's algorithm \cite{johnson73} towards $y \to \infty$ where, again, the badlands function vanishes. The reflection matrix ${\bf R}$ is then identified by matching ${\bf Z}$ to
\be
{\bf \Psi}(y) \simeq \exp \left ( - i {\bf K} y \right ) + \exp \left ( i {\bf K} y \right ) {\bf R},
\ee
where ${\bf K} = {\rm diag} (k_n)$. In particular, we calculate
\be
{\bf R} = \exp \left ( -i {\bf K} y \right ) \left [ \left ( i {\bf K} - {\bf Z} \right )^{-1} \left ( i {\bf K} + {\bf Z} \right ) \right ] \exp \left ( -i {\bf K} y \right ),
\ee
which is independent of $y$ for $y \to \infty$. The reflection coefficients $r_n$ are the entries of $R_{n0}$.

The algorithm is as follows \cite{johnson73}: Let $M$ be the even number of equidistant grid points, $y_m = y_0 + m \Delta y$, where, in our case, $y_0 = 0$ and $\Delta y = y_{\rm end} / M$, and let ${\bf U}_m = {\bf U}(y_m)$. The matrix is initialized with
\be
{\bf Y}_0 = {\bf Z}(0) - \frac{\Delta y}{3} {\bf U}_0,
\ee
and propagated in two steps,
\be
{\bf Y}_{m+1} = \left ( {\bf I} + \Delta y {\bf Y}_m \right )^{-1} {\bf Y}_m - \frac{4 \Delta y}{3} \left ( {\bf I} + \frac{\Delta y^2}{6} {\bf U}_{m +1} \right )^{-1} {\bf U}_{m + 1},
\ee
and
\be \label{eq:evenstep}
{\bf Y}_{m+2} = \left ( {\bf I} + \Delta y {\bf Y}_{m+1} \right )^{-1} {\bf Y}_{m+1} - \frac{2 \Delta y}{3} {\bf U}_{m + 2}.
\ee
Finally, in the last step Eq.~\eqref{eq:evenstep} is replaced by
\be
{\bf Z}(y_{\rm end}) \equiv {\bf Y}_{M} = \left ( {\bf I} + \Delta y {\bf Y}_{M-1} \right )^{-1} {\bf Y}_{M-1} - \frac{\Delta y}{3} {\bf U}_{M}.
\ee
The error decreases as $O(h^4)$ \cite{johnson73}.

% \bibliography{qreflection.bib}

\bibliographystyle{apsrev4-1}

\end{document}